\documentclass[12pt]{article}
\usepackage{t1enc}
\usepackage[utf8]{inputenc}
\usepackage[english]{babel}
\usepackage{amssymb}
\usepackage{amsmath}
\usepackage{amsthm}
\usepackage{amsfonts}
\usepackage{slashed}
\usepackage{bbm}
\usepackage{hyperref}
\usepackage{cite}
\usepackage{authblk}

\date{}

\title{On the positivity preservation of the free Euclidean Dirac equation}
\author{Norbert Barankai}
\affil{\small{MTA-ELTE Theoretical Physics Research Group,\\ E\"{o}tv\"{o}s University,\\ 
P\'{a}zm\'{a}ny P\'{e}ter s\'{e}t\'{a}ny 1/A, H-1117, Budapest, Hungary\\ \vspace{5mm}\texttt{email: barankai@caesar.elte.hu}}}
\begin{document}
\maketitle
\begin{abstract}
We show that the free Euclidean Dirac equation can preserve the positivity of its initial data only in two spacetime dimensions. 
\end{abstract}
\section{Introduction}
Feynman's checkerboard model \cite{FeynmanHibbsStyer} describes the motion of a particle on a line whose inner state determines the direction of that motion: if the particle is in the $+$ ($-$) state, it moves uniformly with a constant velocity $c$ to the right (left). Between two consecutive instants of time separated by the time interval $\Delta t$, the particle can switch its inner state with a complex amplitude proportional to $\mathrm{i}\Delta t$. The continuum limit of the sum-over-histories results in the path integral representation of the propagator of the $1+1$ dimensional Dirac equation \cite{Jacobson1984}, \cite{JacobsonSchulman1984}.

If we replace the complex amplitudes with the probability $\lambda \Delta t$ for the reversals and $1-\lambda \Delta t$ for non-reversals and look on the model as a stochastic process then what we get is the so called persistent random walk \cite{Kac1974}. It has the remarkable property that its continuous time limit is described by an equation whose solution set is - up to a multiplicative transformation - equivalent to that of the $1+1$ dimensional free Euclidean Dirac equation \cite{Gaveau1984a}. Denoting the probability densities corresponding to the spatial distribution of the different states of the particle by $p_{\pm}(x,t)$, the vector $p(x,t)=(p_{+}(x,t),p_{-}(x,t))^T$ satisfies the equation
\begin{equation}
\partial_tp=\lambda(\sigma_1-\mathbbm{1}_2)p - c\sigma_3\partial_x p,
\label{PER}
\end{equation}
where $\sigma_{1,3}$ are Pauli matrices in the standard notation. This equation preserves the non-negativity and normalization, i.e.
\begin{equation*}
0\leq p_{\pm}(x,t)\ \mathrm{and}\ \int_{-\infty}^{\infty}(p_+(x,t)+p_{-}(x,t))\mathrm{d} x=1
\end{equation*}
for all future time instances $t$, provided that they were initially satisfied. The quantity $\psi(x,t)=\mathrm{e}^{\lambda t}p(x,t)$ satisfies the free Euclidean Dirac equation
\begin{equation*}
\partial_t\psi=\lambda \sigma_1\psi - c\sigma_3\partial_x \psi.
\end{equation*}
Thus, the analytic continuation of the parameter $\lambda$ to pure imaginary values by the replacement $\lambda\mapsto mc^2/\mathrm{i}\hbar$ leads to the equation of motion of a massive Dirac particle in two spacetime dimensions in the Weyl representation:
\begin{equation}
\label{WDIRAC}
\mathrm{i}\hbar\partial_t\psi=mc^2\sigma_1\psi - \mathrm{i}\hbar c \sigma_3\partial_x \psi.
\end{equation}

This observation was used \cite{Gaveau1984a} to construct stochastic processes in commutative spaces of operators whose certain expectation values satisfy evolution equations which are closely related to the $1+1$ dimensional Dirac equation and the telegrapher's equation \cite{Goldstein1951} in the same dimension. It was later shown \cite{McKeonOrd1992} that the complex Dirac equation can be obtained without analytic continuation by choosing an appropriate representation of the Clifford algebra $C\ell_{1,1}(\mathbb{R})$ and introducing a random walk on the two dimensional spacetime lattice that enables particles to move backwards in time as well. The $1+1$ dimensional Dirac equation in external field also allows a stochastic interpretation, see \cite{Angelisetal1984}. 

By now, the rigorous foundation of the path integral for the Dirac equation has been elaborated. It has been shown \cite{Ichinose1984a},\cite{Ichinose1984b},\cite{Blanchard1987},\cite{Zastawniak1988} that there exists a countably additive $M_2(\mathbb{C})$ valued measure on the path space $C([0,t],\mathbb{R})$ with which the solution of the Cauchy problem of (\ref{WDIRAC}) with an additional external field can be constructed using Feynman-Kac type theorems \cite{Kac1949}, \cite{Simon1979}. Though the generalization of the method to higher dimensions turned out to be impossible \cite{Zastawniak1989}, path integrals for the Dirac equation exists in momentum \cite{Nakamura2000} and in phase space \cite{Ktitarev1993},\cite{Ichinose2014} for arbitrary dimensions. Furthermore, a Poisson process and an associated time evolution operator can be constructed which generates the solution of the Cauchy problem of the Dirac equation with a stationary external field \cite{Gaveau1984b}. Furthermore, there exists Grassmann valued stochastic processes such that the propagator of the Dirac equation can be written as certain path integrals along the trajectories of such processes \cite{Gaveau1987}. The underlying theory of fermionic Brownian motion was presented in \cite{Rogers1987}, see \cite{Rogers2003} for more details and relations to the Atiyah-Singer index theorem. Recently, some concepts arising from ,,superstatistics'' has been used to derive nonrigorously the relativistic propagation of fermionic fields by mixing together nonrelativistic Brownian paths of different diffusion constants\cite{JizbaScardigli2012},\cite{JizbaScardigli2013},\cite{JizbaScardigli2014}.

On the other hand, the generalization of the persistent random walk to higher spatial dimensions is straightforward. Unfortunately, when the formal continuum limit is taken, the resulting differential equations are unable to reproduce the higher dimensional Dirac equation \cite{Godoyetal1996}, \cite{Bogunaetal1998}. This has been observed several times and - as far as the author of the present paper knows - there is an agreement that the construction of a random-walk-like scenario that could be described by the massive Dirac equation in the continuum limit is impossible or - to say the least - hard to achieve \cite{Dunkeletal2007}, \cite{Plyukhin2010}.

In this paper we shed some light on the reason behind that. If a solution of a PDE can be obtained by - vaguely speaking - measuring the spatial distribution of a  particle moving randomly in space then this PDE must preserve the possible non-negativity and normalization of its initial data. In this case, the solution behaves as a density of a probability measure, inducing a time evolution of such measures. If this happens for every suitable nonnegative and normalized initial data, we say that the PDE preserves positivity. Positivity preservation has been extensively studied for many systems, for example the Fokker-Planck-Kolmogorov equation \cite{BKRS}, the Einstein–Boltzmann system \cite{Lee2013}, the Cattaneo equation \cite{Brasiello2016} or parabolic systems coupled to discrete degrees of freedom \cite{Chabrowski1973} just to name a few - the list could be continued endlessly.  

The result of this paper shows that - beyond $C\ell_{1,1}(\mathbb{R})$ - there is no finite dimensional (faithful) representation of the Clifford algebras $C\ell_{1,d}(\mathbb{R})$ such that the corresponding free Euclidean Dirac equation, the natural generalization of (\ref{PER}) to higher dimensions, could preserve positivity. 

\section{Mathematical preliminaries and notation}
The Clifford algebra $C\ell_{m,n}(\mathbb{R})$ is the unital associative algebra over the reals defined through its generators
\begin{equation*}
C\ell_{m,n}(\mathbb{R})=\langle\mathbbm{1},u_1,\dots,u_{m+n}|\{u_k,u_l\}=2Q_{kl}\mathbbm{1}\rangle,
\end{equation*}
where $\{\cdot,\cdot\}$ denotes the anticommutator of two algebra elements and $Q$ is the diagonal matrix of a real quadratic form with signature $(m,n)$. Only a few is used from the theory of Clifford algebras and those can be found in \cite{trautman2006}. A representation of a Clifford algebra is a faithful, unital algebra homomorphism $R:C\ell_{m,n}(\mathbb{R})\rightarrow \mathrm{M}_S(\mathbb{K})$, where $\mathrm{M}_S(\mathbb{K})$ is the $S$ dimensional matrix algebra over the reals ($\mathbb{K}=\mathbb{R}$) or the complex ($\mathbb{K}=\mathbb{C}$) numbers. In all the representations of $C\ell_{n,0}(\mathbb{R})$, with $1<n$, the images of the generators are traceless matrices \cite{Thaller1993}. Pauli matrices are frequently used in the theory of Clifford algebras, their notation here follows the standard convention:
\begin{equation*}
\sigma_1=\left(\begin{array}{cc} 0 & 1 \\ 1 & 0 \end{array}\right)
\qquad 
\sigma_2=\left(\begin{array}{cc} 0 & -\mathrm{i} \\ \mathrm{i} & 0 \end{array}\right)
\qquad
\sigma_3=\left(\begin{array}{cc} 1 & 0 \\ 0 & -1 \end{array}\right).
\end{equation*}

We use some simple definitions and facts of the theory of finite dimensional linear algebra. All of them can be found in \cite{Minc1988}. A real matrix is nonnegative if all of its entries are nonnegative. A matrix $M$ is called reducible if there is a permutation matrix $\Pi$ such that $\Pi M\Pi^{-1}$ is a block upper-triangular matrix. The matrix $M$ is irreducible if such a permutation matrix cannot be found. The Perron-Frobenius theorem states, among others, that the spectral radius of the irreducible nonnegative $M$ is a positive real number and it is always a simple eigenvalue of $M$. Furthermore, the eigenspace corresponding to the spectral radius contains a vector with strictly positive entries. Throughout the paper $\mathbbm{1}_S$ denotes the $S$ dimensional unit matrix and $\mathbf{1}_S\in\mathbb{R}^S$ denotes the vector $S^{-1/2}(1,1,\dots,1)^T$. The standard basis vectors of $\mathbb{R}^S$ are denoted by $\mathbf{e}^{(S)}_m$: the subscript here runs from $1$ to $S$.

If $u:\mathbb{R}^d\rightarrow \mathbb{C}$ is absolute integrable, its Fourier transform is denoted by $\Phi[u]$:
\begin{equation*}
\Phi[u](\mathbf{k})=\int_{\mathbb{R}^d}u(\mathbf{x})\mathrm{e}^{\mathrm{i}\mathbf{k}\cdot\mathbf{r}}\mathrm{d}^{d}\mathbf{x}.
\end{equation*}
We follow the same notation even if $u$ is an array of functions: in that case $\Phi[u]$ denotes the array containing the Fourier transforms of the entries of $u$. Dependence of some quantity $u$ on a time parameter is usually emphasized by writing $u(t)$ not forgetting that $u$ may depend on several other variables. If $u$ is an array, then any appearance of expressions like $0\leq u$ is understood as an element-wise inequality.

The natural function space of initial data of the Dirac equation is the space of $S$ dimensional vectors whose entries are members of the first Sobolev space of $L^{2}(\mathbb{R}^d,\mathbb{C})$ \cite{Thaller1993}. This function space is denoted by $H^{1}(\mathbb{R}^d)^{\oplus S}$, $S$ is the dimension of the representation of $C\ell_{1,d}(\mathbb{R})$. 

Let $\mu$ be the product of the counting measure on $\{1,\dots,S\}$ and the Lebesgue measure on $\mathbb{R}^d$. It is a measure on the Borel sigma algebra of the set $\Omega(S,d)=\{1,\dots,S\}\times \mathbb{R}^d$. If $\mathbb{P}$ is a probability measure on $\Omega(S,d)$, it is always meant to be absolutely continuous with respect to $\mu$. The Radon-Nikodym derivative of such a measure is called probability density. If $f:\mathbb{R}^d\rightarrow \mathbb{R}$ is a nonnegative function, its Fourier transform is positive definite, i.e.~for any finite set $\{\mathbf{k}_1,\dots,\mathbf{k}_N\}\subset\mathbb{R}^d$  the matrix $F\in\mathrm{M}_N(\mathbb{C})$ defined by its entries $(F)_{ab}=f(\mathbf{k}_a-\mathbf{k}_b)$ is positive definite. On the contrary, every continuous, positive definite complex valued function over $\mathbb{R}^d$ is the Fourier transform of a nonnegative function, integrable on the whole $\mathbb{R}^d$. See Bochner's theorem that gives a complete characterization of characteristic functions of probability measures of locally compact abelian groups \cite{HewittRoss1963},\cite{Kawata2014}.

\section{Positivity preservation}
The free Euclidean Dirac equation in $d+1$ dimensions with a mass term has the dimensionless form 
\begin{equation}
\sum_{\mu=0}^{d}\gamma_\mu\partial_\mu \psi =\psi,
\label{ODIRAC}
\end{equation}
where $\{\gamma_\mu\}_{\mu=0}^{d}$ is the hermitian image of a set of generators of $C\ell_{1,d}(\mathbb{R})$ in one of its $S$ dimensional real representations satisfying the relations $\{\gamma_\mu,\gamma_\nu\}=2\eta_{\mu\nu}\mathbbm{1}_S$ with $\eta=\mathrm{diag}(1,-1,\dots,-1)$. The spinor $\psi:\mathbb{R}^{d}\times\mathbb{R}\rightarrow\mathbb{\mathbb{C}}^{S}$ describes the (non-unitary) propagation of a free spin-$1/2$ particle in $d+1$ dimensional spacetime. If $\psi$ is a solution of the equation, then $p(t)=\mathrm{e}^{-\alpha t}\gamma_0\psi(t)$ satisfies 
\begin{equation}
\partial_t p = -\alpha p+e_0p-\sum_{\mu=1}^{d} e_\mu\partial_\mu  p,
\label{DIRAC}
\end{equation}
where $e_0=\gamma_0$ and $e_\mu=\gamma_\mu\gamma_0$ for any other $1\leq \mu\leq d$. From now on, since the solution space of (\ref{ODIRAC}) and (\ref{DIRAC}) are equivalent we call (\ref{DIRAC}) the massive (Euclidean) Dirac equation. The matrices $\{e_\mu\}_{\mu=0}^{d}$ satisfy the anticommutation relations $\{e_\mu,e_{\nu}\}=2\delta_{\mu\nu}\mathbbm{1}_S$, thus they are images of generators of the Clifford algebra $C\ell_{d+1,0}(\mathbb{R})$ in one of its representations. We are interested in those representations where these matrices are real and symmetric.  

If $p(t)$ is a solution of (\ref{DIRAC}) for all $0\leq t$, the time evolution of its (spatial) Fourier transform $\Phi[p(t)]$ is governed by the linear autonomous system
\begin{equation*}
\partial_t \Phi[p(t)](\mathbf{k}) = -\alpha \Phi[p(t)](\mathbf{k})+e_0\Phi[p(t)](\mathbf{k})+\mathrm{i}\sum_{\mu=1}^{d} k_\mu e_\mu\Phi[p(t)](\mathbf{k})
\end{equation*}
for all $\mathbf{k}=(k_1,\dots,k_d)^T\in\mathbb{R}^d$. Using the anticommutation relations of the Clifford algebra $C\ell_{d+1,0}(\mathbb{R})$, the time evolution $\Phi[p(t)](\mathbf{k})=\mathcal{D}(\mathbf{k},t)\Phi[p(0)](\mathbf{k})$ can be computed explicitly: 
\begin{equation}
\mathcal{D}(\mathbf{k},t) = \mathrm{e}^{-\alpha t}\cosh(\beta(\mathbf{k})t)\mathbbm{1}_S+\mathrm{e}^{-\alpha t}\beta^{-1}(\mathbf{k})\sinh(\beta(\mathbf{k})t)\left(e_0+\mathrm{i}\sum_{\mu=1}^{d} k_\mu e_\mu \right)
\label{D}
\end{equation}
where $\beta(\mathbf{k})=\sqrt{1-2\mathbf{k}^2}$. The inverse Fourier transform of $\mathcal{D}(t)\Phi[p(0)]$ gives the unique solution of (\ref{DIRAC}) with initial data belonging to the Sobolev space $H^{1}({\mathbb{R}^d})^{\oplus S}$. Motivated by the fact that the generators $\{e_{\mu}\}_{\mu=0}^{d}$ are real matrices, we split $\mathbb{C}^S$ to real and imaginary subspaces: $\mathbb{C}^S\simeq \mathbb{R}^S\oplus \mathbb{R}^S$. Since $\beta(\mathbf{k})$ has purely real (imaginary) values for any $\mathbf{k}\in\mathbb{R}^d$ whose norm is less or equal to (greater than)  $2^{-1/2}$, the functions $t\mapsto \cosh(\beta(\mathbf{k})t)$  and $t\mapsto \beta^{-1}(\mathbf{k})\sinh(\beta(\mathbf{k})t)$ are purely real for all $\mathbf{k}\in\mathbbm{R}^d$. Thus, the decomposition of $\mathcal{D}(t)$ with respect to the above splitting is 
\begin{align}
\mathcal{D}_{00}(\mathbf{k},t)&=\mathcal{D}_{11}(\mathbf{k},t)= \mathrm{e}^{-\alpha t}\left(\cosh(\beta(\mathbf{k})t)\mathbbm{1}_S+\beta^{-1}(\mathbf{k})\sinh(\beta(\mathbf{k})t)e_0\right)\nonumber\\
\mathcal{D}_{01}(\mathbf{k},t)&=-\mathrm{e}^{-\alpha t}\beta^{-1}(\mathbf{k})\sinh(\beta(\mathbf{k})t)\sum_{\mu=1}^{d} k_\mu e_\mu\nonumber \\
\mathcal{D}_{10}(\mathbf{k},t)&=+\mathrm{e}^{-\alpha t}\beta^{-1}(\mathbf{k})\sinh(\beta(\mathbf{k})t)\sum_{\mu=1}^{d} k_\mu e_\mu.
\end{align}
where the indices $0$ and $1$ correspond to real and imaginary subspaces, respectively. 

Let $\mathbb{P}_0$ be a probability measure on $\Omega(S,d)$. Since the Radon-Nikodym derivative of $\mathbb{P}_0$ with respect to $\mu$ is unique up to a measure zero subset of $\Omega(S,d)$, then either all of its probability densities are members of $H^{1}(\mathbb{R}^d)^{\oplus S}$ or none of them. Assume the former and let $p_0$ be a probability density of $\mathbb{P}_0$. Set the initial condition of (\ref{DIRAC}) to be equal to $p_0$ and let $p(t)$ be the corresponding solution as described above. Define $\mathbb{P}(t)$ by the assignment $\Omega(S,d)\supseteq A\mapsto \int_A p(t)d\mu$. Note that $\mathbb{P}(t)$ does not depend on the particular choice of the representative $p_0$ in the corresponding equivalence class in $H^{1}(\mathbb{R}^d)^{\oplus S}$. We say that \par
\vspace{5mm}
\noindent
\textbf{Definition.} \textit{The Dirac equation (\ref{DIRAC}) preserves positivity if for every probability measure $\mathbb{P}_0$ whose densities are in $H^{1}({\mathbb{R}^d})^{\oplus S}$, $\mathbb{P}(t)$ is a probability measure on $\Omega(S,d)$ for all $0\leq t$.} 
\vspace{5mm}
\par
\noindent
The result of our paper is the following theorem.\par
\vspace{5mm}
\noindent
\textbf{Theorem.} \textit{Let $\{e_{\mu}\}_{\mu=0}^{d}$ be the image of a set of generators of the Clifford algebra $C\ell_{d+1,0}(\mathbb{R}^d)$ in one of its representations on $\mathrm{M}_S(\mathbb{R})$. Assume that these matrices are symmetric. Then, the massive Dirac equation 
\begin{equation*}
\partial_t p = -\alpha p+e_0p-\sum_{\mu=1}^{d}e_\mu\partial_\mu p
\end{equation*}
preserves positivity if and only if the following conditions are simultaneously satisfied:
\begin{enumerate}
\item $d=1$,
\item there exists a permutation matrix $\Pi$ and a positive integer $0<m$ such that
\begin{equation*}
\Pi e_0\Pi^{-1}=\sigma_1^{\oplus m} \qquad  \Pi e_1\Pi^{-1}=\sigma_3^{\oplus m},
\end{equation*}
\item  and $\alpha=1$.
\end{enumerate}
}
\noindent
\textbf{Proof.} $\Rightarrow$: Let $p(t)$ be the solution of the massive Dirac equation with the initial condition $p(0)=p_0\in H^{1}(\mathbb{R}^d)^{\oplus S}$, where $p_0$ is a density of a probability measure $\mathbb{P}_0$ on $\Omega(S,d)$. Since the Dirac equation preserves positivity, $p(t)$ is (up to a measure zero subset of $\Omega(S,d)$) a nonnegative function, so the Fourier transform $\Phi[p(t)]$ must be an array of positive definite functions through the whole time evolution. In the following real and imaginary parts of $\Phi[p(t)]$ are denoted by $\varphi(t)$ and $\chi(t)$, respectively. 

Amongst the many, two necessary conditions of positive definiteness and normalization are the following. Firstly, $\varphi(t)$ must satisfy the inequality:
\begin{equation}
0\leq\int_{\mathbb{R}^d}(1+\cos(\mathbf{k}\cdot\mathbf{x}))p(\mathbf{x},t)\mathrm{d} ^d\mathbf{x}=\varphi(\mathbf{0},t)+\varphi(\mathbf{k},t)
\label{INEQ}
\end{equation}
for all positive $t$. Secondly, provided that the zero mode of $\varphi(t)$ is a measure on the discrete set $\{1,\dots,S\}$, it is nonnegative and normalized:
\begin{equation}
1=\sum_{p=1}^{S}\varphi_p(\mathbf{0},t)\qquad 0\leq \varphi(\mathbf{0},t)
\label{ZERO}
\end{equation}
for all positive $t$. 

The time evolution of $\varphi(\mathbf{0},t)$ is governed by $\mathcal{D}_{00}(\mathbf{0},t)$. The conditions in (\ref{ZERO}) tells that this matrix must be a right stochastic matrix for all $0\leq t$: its entries are nonnegative and in each column they sum up to one. The concrete form 
\begin{equation*}
\mathcal{D}_{00}(\mathbf{0},t)=\mathrm{e}^{-\alpha t}(\cosh(t)\mathbbm{1}_S+\sinh(t)e_0)
\end{equation*}
and the non-negativity condition constrain the possible range of the diagonal and off-diagonal entries of $e_0$:
\begin{equation}
-1\leq (e_0)_{pp}\qquad  0\leq (e_0)_{pq} \quad (p\neq q) 
\label{INEQ_ZERO}
\end{equation}
and impose the equality
\begin{equation}
\cosh(t)+\sinh(t)\sum_{p=1}^{S}(e_0)_{pq}=\mathrm{e}^{\alpha t}
\label{EQ_ZERO}
\end{equation}
for all $1\leq p, q\leq S$ and $0\leq t$. In the following, we will consider two separate cases depending on whether $e_0$ is irreducible or not.\par
\noindent\textit{The matrix $e_0$ is irreducible.} As a consequence of (\ref{INEQ_ZERO}) the projection $P^{+}_0=(1+e_0)/2$ is nonnegative. The irreducibility of $e_0$ implies the irreducibility of $P^{+}_0$ also. By the Perron-Frobenius theorem, its largest eigenvalue is simple and the corresponding eigenspace contains a right eigenvector with strictly positive entries. That is, $P^{+}_0$ is a rank one projection: $P^{+}_0=uu^T$, where $u\in\mathbb{R}^S$ has unit length and its entries are real, non-vanishing and share the same sign. As a consequence, $1$ is a simple eigenvalue of the generator $e_0=2uu^T-1$ and its other eigenvalue, $-1$ is degenerate with multiplicity $S-1$. Since $\mathrm{Tr} (e_0)=0$, the representation must be two dimensional and the only Clifford algebra among $C\ell_{d,0}(\mathbb{R})$ which has at least two generators and a (faithful) representation in $\mathrm{M}_2(\mathbb{R})$ is $C\ell_{2,0}(\mathbb{R})$. Substitution of $e_0=2uu^T-1$ into (\ref{EQ_ZERO}) gives
\begin{equation}
\mathrm{e}^{-t}+ (\mathrm{e}^{t}-\mathrm{e}^{-t})u_q\sum_{p=1}^{2}u_p=\mathrm{e}^{\alpha t}.
\label{ALPHA_EQ}
\end{equation}
As all entries of $u$ are strictly positive or strictly negative, $\sum_{p=1}^{2}u_p$ does not vanish, so (\ref{ALPHA_EQ}) holds for all $q\in\{1,2\}$ and all $0\leq t$ if and only if $u$ is equal to $\pm\mathbf{1}_2$ and $\alpha$ is equal to one. Since $e_0=2\mathbf{1}^{}_2\mathbf{1}^T_2-\mathbbm{1}_2=\sigma_1$, $e_1$ must be equal to $\sigma_3$ or $-\sigma_3$. Choosing $m=1$ and $\Pi=\mathbbm{1}_2$ in the first or $\Pi=\sigma_1$ in the second case, respectively, we arrive to the desired result. \par
\noindent\textit{The matrix $e_0$ is reducible.} If $e_0$ is symmetric and reducible, there must be a permutation matrix $\Pi$ such that both $\Pi e_0\Pi^{-1}$ and $\Pi\mathcal{D}_{00}(\mathbf{0},t)\Pi^{-1}$ are block diagonal matrices whose blocks are irreducible. Furthermore, the blocks of $\Pi\mathcal{D}_{00}(\mathbf{0},t)\Pi^{-1}$ are right stochastic matrices. The form of the conditions of (\ref{INEQ_ZERO}) and (\ref{EQ_ZERO}) are invariant under similarity transformations by permutation matrices enabling us to apply the Perron-Frobenius theorem to each of the blocks of $\Pi e_0\Pi^{-1}$ and - with the help of $\mathrm{Tr} (e_0)=\mathrm{Tr} (\Pi e_0\Pi^{-1})=0$ - conclude that $\alpha=1$ and there must be a positive integer $0<m$ such that $\Pi e_0\Pi^{-1}=\sigma_1^{\oplus m}$. Let $f\in H^{1}(\mathbb{R}^d)$ be a probability density on $\mathbb{R}^d$ having even parity and a strictly positive real part of its Fourier transform. For example, the multivariate Cauchy distribution $f(\mathbf{x})=\pi^{-d}\prod_{\mu=1}^{d}(1+x^2_\mu)^{-1}$ satisfies these conditions. Set the initial condition to $p_0 =f_{\mathbf{a}}\Pi^{-1}\mathbf{e}^{(S)}_q$, where $f_{\mathbf{a}}(\mathbf{x})=f(\mathbf{x}-\mathbf{a})$. The Fourier transform of the initial data is 
\begin{eqnarray}
\varphi(\mathbf{k},0)=\cos(\mathbf{k}\cdot\mathbf{a})g(\mathbf{k})\Pi^{-1}\mathbf{e}^{(S)}_q,\nonumber\\
\chi(\mathbf{k},0)=\sin(\mathbf{k}\cdot\mathbf{a})g(\mathbf{k})\Pi^{-1}\mathbf{e}^{(S)}_q,
\end{eqnarray}
where $g$ is the (strictly positive) Fourier transform of $f$. The entry-wise inequalities in (\ref{INEQ}) are invariant under the action of any permutation matrix so we can write  
\begin{equation}
0\leq \widehat{\mathcal{D}}_{00}(\mathbf{0},t)\Pi\varphi(\mathbf{0},0)+\widehat{\mathcal{D}}_{00}(\mathbf{k},t)\Pi\varphi(\mathbf{k},0)+\widehat{\mathcal{D}}_{01}(\mathbf{k},t)\Pi\chi(\mathbf{k},0),
\label{DETAILED_INEQ}
\end{equation} 
where $\widehat{\mathcal{D}}_{ab}(t)$ denotes $\Pi\mathcal{D}_{ab}(t)\Pi^{-1}$. Let $\mathbf{k}=k\mathbf{e}^{(d)}_\nu$ and $\mathbf{a}=a\mathbf{e}^{(d)}_\nu$. Then, simplifying (\ref{DETAILED_INEQ}) by dividing both of its sides with $\mathrm{e}^{-t}$ gives
\begin{equation}
0\leq \sinh(t)(\widehat{e}_0)_{pq}+B_\nu(k,a)(\widehat{e}_0)_{pq}+C_\nu(k,a)k(\widehat{e}_\nu)_{pq},
\label{AUX_INEQ}
\end{equation}
whenever the indices $p$ and $q$ are nonequal. The functions $B_\nu$ and $C_\nu$ are defined as
\begin{eqnarray}
B_\nu(k,a)=\left(1-2k^2\right)^{-1/2}g\left(k\mathbf{e}^{(d)}_\nu\right)\cos(ka)\sinh\left(t\left(1-2k^2\right)^{1/2}\right),\nonumber\\
C_\nu(k,a)=-\left(1-2k^2\right)^{-1/2}g\left(k\mathbf{e}^{(d)}_\nu\right)\sin(ka)\sinh\left(t\left(1-2k^2\right)^{1/2}\right).\label{AUX_FUNC}
\end{eqnarray}
and $\widehat{e}_\mu=\Pi e_\mu\Pi^{-1}$. Assume a vanishing $(\widehat{e}_0)_{pq}$ (which is equivalent to the assumption that the off-diagonal entry $(\widehat{e}_0)_{pq}$ lies in an off-diagonal block of $\widehat{e}_0$), pick up a nonzero $K$ in $(2^{-1/2},+2^{-1/2})$ and set $a=\pi/2K$, then the sudden substitution to (\ref{AUX_FUNC}) gives $\sin(ka)=1$ and $\cos(ka)=$, so (\ref{AUX_INEQ}) has the form
\begin{equation}
0\leq -g(K\mathbf{e}^{(d)}_\nu)\frac{\sinh(t\sqrt{1-2K^2})}{\sqrt{1-2K^2}}K(\widehat{e}_\nu)_{pq}
\label{F_INEQ}
\end{equation}
must hold for all $0\leq t$. But this inequality can be satisfied if and only if $(e_\nu)_{pq}$ vanishes: in any other case, the r.h.s~of (\ref{F_INEQ}) is an odd function of $K\in (-2^{-1/2},+2^{-1/2})$. As a consequence, the spatial generators $\widehat{e}_\mu$ must be all block-diagonal matrices with each diagonal block belonging to $\mathrm{M}_2(\mathbb{R})$. In that case, the Clifford algebra relations between the appropriate blocks can be satisfied only if $d$ is equal to one and $\Pi e_1\Pi^{-1}$ is a block diagonal matrix whose non-vanishing blocks are equal to $\sigma_3$ or $-\sigma_3$. Again, a permutation matrix $C$ can always be chosen such that
\begin{equation*}
(C\Pi) e_1(C\Pi )^{-1}=\sigma_3^{\oplus m}.
\end{equation*}
Such a $C$ leaves $e_0$ invariant, that is 
\begin{equation*}
(C\Pi) e_0(C\Pi )^{-1}=\sigma_1^{\oplus m}
\end{equation*}
which completes the proof in the forward direction.\par
\noindent
$\Leftarrow$: It is enough to consider the $m=1$ case. We have to show that the massive Dirac equation  
\begin{equation}
\partial_t p =-p+\sigma_1p-\sigma _3\partial_xp 
\label{DIRAC1+1}
\end{equation}
preserves positivity. We will exploit the fact that (\ref{DIRAC1+1}) is the ballistic scaling limit of the persistent random walk \cite{Gaveau1984a}. Fixing $k\in\mathbb{R}$, the application of the Lie-Trotter formula to calculate (\ref{D}) gives 
\begin{equation*}
\mathcal{D}(k,t)=\lim_{N\rightarrow\infty}(V(t/N)S(k,t/N))^N,
\end{equation*}
where 
\begin{equation*}
V(\tau)=\frac{1}{2}\left(\begin{array}{cc} 1+\mathrm{e}^{-2\tau} & 1-\mathrm{e}^{-2\tau} \\ 1-\mathrm{e}^{-2\tau} & 1+\mathrm{e}^{-2\tau} \end{array}\right)
\end{equation*}
and 
\begin{equation*}
S(k,\tau)=\left(\begin{array}{cc} \mathrm{e}^{\mathrm{i} k \tau} & 0 \\ 0 & \mathrm{e}^{-\mathrm{i} k \tau} \end{array}\right).
\end{equation*}
Under the inverse Fourier transform, $S(\tau)$ maps to $T(\tau)$ defined by 
\begin{equation*}
T(\tau)\left(\begin{array}{c} f_1(x) \\ f_2(x)\end{array}\right)=\left(\begin{array}{c} f_1(x-\tau) \\ f_2(x+\tau)\end{array}\right)
\end{equation*}
and $V(\tau)$ remains invariant. Both $V(\tau)$ and $T(\tau)$ leaves $H^{1}(\mathbb{R})^{\oplus 2}$ invariant so the functions $p^{(N)}(t):\mathbb{R}\rightarrow \mathbb{R}$ defined by 
\begin{equation*}
p^{(N)}(t)=(V(t/N)T(t/N))^Np_0
\end{equation*}
are members of $H^{1}(\mathbb{R})^{\oplus 2}$ and approximate the solution
\begin{equation*}
p(x,t)=\frac{1}{2\pi}\int_{\mathbb{R}}\mathcal{D}(k,t)\Phi[p_0](k)\mathrm{e}^{-\mathrm{i} kx}\mathrm{d} k\nonumber
\end{equation*}
of (\ref{DIRAC1+1}) whenever $p_0\in H^{1}(\mathbb{R})^{\oplus 2}$. Furthermore, both $V(\tau)$ and $T(\tau)$ preserve nonnegativity and normalization for positive $\tau$, so if $p_0$ shares these properties, $p(t)$ can be approximated by nonnegative and normalized functions with arbitrary precision, thus (up to a measure zero subset) $p(t)$ is nonnegative and normalized for all $0\leq t$. $\blacksquare$

\section{Conclusion}
In the previous section we proved that the Dirac equation can sustain positivity and normalization of initial data only in the $1+1$ dimensional case. Even in two dimensional spacetime, the form of the equation is unique: it is at most an assembly of independent particles performing persistent random walk in the line and described individually by equation (\ref{PER}). This means that the Dirac equation cannot provide the time evolution of a stochastic process directly in spatial dimensions greater than one.

\section*{Acknowledgements}
The author would like to thank Tam\'{a}s Temesv\'{a}ri and P\'{e}ter Vecserny\'{e}s for discussions.

\end{document}